\documentclass{aa}

\usepackage{amsmath}           % AMS extended math package
\usepackage{dcolumn}           % column alignment enhanc
\usepackage{graphicx}          % EPS inclusion
\usepackage{txfonts}           % fonts
\usepackage{mathrsfs}          % Raph's Smith Formal Script
\usepackage{rotating}          % rotates pages 
\usepackage{subfigure}         % multiple figures
\usepackage{lscape}            % landscape pages
\usepackage{afterpage}         % improves control float objects
\usepackage{xspace}            % gentle spacing after macro
\usepackage{natbib}            % package for citations
\usepackage{hyperref}          % link reference
\usepackage{nicefrac}          % oblique fraction
\usepackage{amsfonts}          % AMS extended fonts
\usepackage{mathrsfs}          % additional math operator
\usepackage{textcomp}
\usepackage{chemformula}       % chemical formula
\let\ce\ch

\providecommand*{\arcmin}{\ensuremath{^{\prime}}\xspace}
\providecommand*{\arcsec}{\ensuremath{^{\prime\!\prime}}\xspace}
\providecommand*{\wn}{\,cm$^{-1}$\xspace}

\newcommand{\mcl}[3]{\multicolumn{#1}{#2}{#3}}
\newcommand{\mrm}[1]{\ensuremath{\mathrm{#1}}}

\newcolumntype{.}{D{.}{.}{-1}}
\newcolumntype{d}[1]{D{.}{.}{#1}}

\begin{document}

	\title{Far-infrared laboratory spectroscopy of aminoacetonitrile and first interstellar detection of its vibrationally excited transitions \thanks{The list of assigned transitions is only available in electronic form at the CDS via anonymous ftp to cdsarc.u-strasbg.fr (130.79.128.5) or via http://cdsweb.u-strasbg.fr/cgi-bin/qcat?J/A+A/}}

	\author{M.~Melosso\inst{1}
		\and
		A. Belloche\inst{2}
		\and
		M.-A.~Martin-Drumel\inst{3}
		\and
		O.~Pirali\inst{3}\fnmsep\inst{4}
		\and
		F.~Tamassia\inst{5}
		\and
		L.~Bizzocchi\inst{6}
		\and
		R.T.~Garrod\inst{7}\fnmsep\inst{8}
		\and
		H.S.P.~M\"uller\inst{9}
		\and
		K.M.~Menten\inst{2}
		\and
		L.~Dore\inst{1}
		\and
		C.~Puzzarini\inst{1}
	}
	
	\institute{Dipartimento di Chimica ``Giacomo Ciamician'', Universit\`a di Bologna, via F.~Selmi~2, 40126 Bologna, Italy \\
		\email{mattia.melosso2@unibo.it}
		\and
		Max-Planck-Institut f\"ur Radioastronomie, Auf dem H\"ugel 69, 53121 Bonn, Germany
		\and
		Universit\'e Paris-Saclay, CNRS, Institut des Sciences Mol\'eculaires d'Orsay, 91405 Orsay Cedex, France
		\and
		SOLEIL Synchrotron, AILES beamline, l'Orme des Merisiers, Saint-Aubin, 91190 Gif-sur-Yvette, France
		\and
		Dipartimento di Chimica Industriale ``Toso Montanari'', Universit\`a di Bologna, viale del Risorgimento~4, 40136 Bologna, Italy
		\and
		Center for Astrochemical Studies, Max-Planck-Institut f\"ur extraterrestrische Physik, Gie\ss enbachstr.~1, 85748 Garching, Germany
		\and
		Department of Chemistry, University of Virginia, Charlottesville, VA 22904, USA
		\and
		Department of Astronomy, University of Virginia, Charlottesville, VA 22904, USA 
		\and
		I. Physikalisches Institut, Universit\"at zu K\"oln, Z\"ulpicher Str. 77, 50937, K\"oln, Germany
	}
	
	\date{Received --; accepted --}

	\abstract
	% context heading (optional)
	{Aminoacetonitrile, a molecule detected in the interstellar medium only towards the star-forming region Sagittarius B2 (Sgr~B2) thus far, is considered an important prebiotic species; in particular it is a possible precursor of the simplest amino acid glycine. To date, observations were limited to ground state emission lines, whereas transitions from within vibrationally excited states remained undetected.}
	% aims heading (mandatory)
	{We wanted to accurately determine the energies of the low-lying vibrational states of aminoacetonitrile, which are expected to be populated in Sgr~B2(N1), the main hot core of Sgr~B2(N). This step is fundamental in order to properly evaluate the vibration-rotation partition function of aminoacetonitrile as well as the line strengths of the rotational transitions of its vibrationally excited states. This is necessary to derive accurate column densities and secure the identification of these transitions in astronomical spectra.}
	% methods heading (mandatory)
	{The far-infrared ro-vibrational spectrum of aminoacetonitrile has been recorded in absorption against a synchrotron source of continuum emission. Three bands, corresponding to the lowest vibrational modes of aminoacetonitrile, were observed in the frequency region below 500\,\wn. The combined analysis of ro-vibrational and pure rotational data allowed us to prepare new spectral line catalogs for all the states under investigation.
	We used the imaging spectral line survey ReMoCA performed with ALMA to search for vibrationally excited aminoacetonitrile toward Sgr~B2(N1). The astronomical spectra were analyzed under the local thermodynamic equilibrium (LTE) approximation.}
	% results heading (mandatory)
	{Almost 11\,000 lines have been assigned during the analysis of the laboratory spectrum of aminoacetonitrile, thanks to which the vibrational energies of the $\varv_{11}=1$, $\varv_{18}=1$, and $\varv_{17}=1$ states have been determined. The whole dataset, which includes high $J$ and $K_a$ transitions, is well reproduced within the experimental accuracy. Reliable spectral predictions of pure rotational lines can now be produced up to the THz region.
	On the basis of these spectroscopic predictions, we report the interstellar detection of aminoacetonitrile in its $\varv_{11}=1$ and $\varv_{18}=1$ vibrational states toward Sgr~B2(N1) in addition to emission in its vibrational ground state. The intensities of the identified $\varv_{11}=1$ and $\varv_{18}=1$ lines are consistent with the detected $\varv=0$ lines under LTE at a temperature of 200~K for an aminoacetonitrile column density of $1.1 \times 10^{17}$~cm$^{-2}$.}
	% conclusions heading (optional)
	{This work shows the strong interplay between laboratory spectroscopy exploiting (sub)millimeter and synchrotron far-infrared techniques, and observational spectral surveys to detect complex organic molecules in space and quantify their abundances.}
	
	
	\titlerunning{Hot aminoacetonitrile in Sgr~B2}
	
	\maketitle
	
	%-----------------------------------------------------------------
	
	\section{Introduction}
	\label{s:intro}
	
	Sagittarius~B2 (Sgr~B2) is one of the most prominent star forming regions in our Galaxy. This large molecular cloud complex is located at a projected distance of $\sim$100~pc from the center of our Galaxy, at a distance of 8.2~kpc from the Sun \citep[][]{Reid19}. This complex harbors two main protoclusters, Sgr~B2(N) and Sgr~B2(M), which are both forming high mass stars. Sgr~B2 in general, and Sgr~B2(N) in particular, are well known for their rich chemistry, revealed by the detection of numerous complex organic molecules reported over the past five decades (see, e.g., \citeauthor{Menten04} \citeyear{Menten04} for a summary and a list of early detections, and \citeauthor{McGuire18} \citeyear{McGuire18} for a complete census of interstellar molecules). Because of their large number of degrees of freedom leading to large partition functions, complex organic molecules (COMs), i.e. carbon-bearing molecules with at least six atoms \citep[][]{Herbst09}, emit numerous 
	rotational lines in the radio, millimeter, and submillimeter wavelength ranges. These spectral lines are generally weak and the presence of many COMs in sources with a rich chemistry leads to spectra that contain forests of lines, sometimes even reaching the spectral confusion limit. 	
	In this context, unbiased spectral line surveys that cover large frequency ranges are the best tools to identify COMs in the interstellar medium and have indeed led to many of the first interstellar detections mentioned above. In particular, Sgr~B2 has been the target of several spectral line surveys carried out with single-dish radio and millimeter wavelength telescopes \citep[e.g.,][]{Cummins86,Nummelin98,Belloche13,Remijan13}. More recently, some of us started a sensitive imaging spectral line survey of Sgr~B2(N) at high angular resolution and sensitivity with the Atacama Large Millimeter/submillimeter Array (ALMA). The first incarnation of this survey was performed during the first two observation cycles of ALMA and was called EMoCA, which stands for exploring molecular complexity with ALMA \citep[][]{Belloche16}. Among other results, this survey led to the first interstellar detection of a branched alkyl molecule, iso-propyl cyanide \citep[i-\ce{C3H7CN},][]{Belloche14}, and a tentative detection of N-methylformamide \citep[\ce{CH3NHCHO},][]{Belloche17}. The second, more recent incarnation of the survey was performed at even higher angular resolution with ALMA during its Cycle 4 and was called ReMoCA, which simply stands for re-exploring molecular complexity with ALMA. This survey is currently being analyzed and has already led to the first unambiguous interstellar detection of urea and the confirmation of the interstellar detection of N-methylformamide \citep[][]{Belloche19}.
	In this work, we take advantage of the ReMoCA survey to explore further the spectral signatures of aminoacetonitrile (\ce{NH2CH2CN}) in the interstellar medium. The first interstellar detection of this COM in its vibrational ground state was reported toward Sgr~B2(N) by \citet{Belloche08} on the basis of a spectral line survey \citep[][]{Belloche13} performed with the 30~m telescope of the Institut de Radioastronomie Millim\'etrique (IRAM). In the same study, the single-dish identification of aminoacetonitrile was confirmed by the detection of a few transitions at higher angular resolution with the IRAM Plateau de Bure interferometer (PdBI) and the Australia Telescope Compact Array (ATCA). The beam of the 30~m telescope, $\sim$25$\arcsec$ at 100~GHz, enclosed several hot molecular cores of Sgr~B2(N), including the main ones Sgr~B2(N1) and Sgr~B2(N2), but the identified emission of aminoacetonitrile was dominated by the main hot core Sgr~B2(N1) at a velocity of 64~km~s$^{-1}$. The PdBI and ATCA interferometric maps revealed emission of aminoacetonitrile toward Sgr~B2(N1) only. Later, \citet{Richard18} reported the detection of aminoacetonitrile in its vibrational ground state also toward the secondary hot core Sgr~B2(N2) on the basis of the EMoCA survey.
	In hot cores, many COMs were detected not only through rotational transitions in their vibrational ground state but also through rotational transitions that belong to some of their vibrationally excited states \citep[see, e.g.,][]{Belloche13,Belloche19,daly2013laboratory,mueller2016laboratory,Bizzocchi17}. Rotational emission of vibrationally excited aminoacetonitrile was searched for toward Sgr~B2(N2) with EMoCA, but was not detected, as quoted in \citet{degli2017millimeter}.
	Given that the secure identification of new, low-abundance COMs in line-rich interstellar spectra relies on carefully taking account of blends with other species, it is critical to characterize in the laboratory the rotational spectrum of COMs not only in their vibrational ground state but also in their vibrationally excited states.
	
	Moreover, an accurate evaluation of molecular column densities from astronomical observations depends, among other quantities, on the value of the partition function $Q$. Because the value of $Q$ can be greatly affected by low-lying vibrational excited states, it is fundamental to have an accurate knowledge of the energy of ro-vibrational levels.
	Despite the large number of spectroscopic studies focused on the rotational spectra of aminoacetonitrile in the ground state \citep{macdonald1972microwave,pickett1973microwave,brown1977quadrupole,bogey1990millemeter,motoki2013submillimeter} and low-lying excited states \citep{kolesnikova2017rotational,degli2017millimeter}, very little is known about its vibrational spectrum.
	The infrared spectrum of aminoacetonitrile has been investigated only at low resolution, either in the gas phase \citep{bak1975vibrational} or in argon matrix \citep{bernstein2004infrared}. In the latter work, theoretical calculations using the density functional theory (DFT) were also carried out.
	However, even combining the available information, the energy and assignment of all vibrational states remain doubtful.
	In order to shed light on the ro-vibrational manifold of aminoacetonitrile and to solve the residual discrepancies among previous studies, we have explored the far-infrared (FIR) spectrum of aminoacetonitrile using a synchrotron-based experiment.
	Our ultimate goal was to determine the vibrational energies of the low-lying excited states of aminoacetonitrile, which are likely to be observable in the hot core Sgr~B2(N1).
	
	In this work, we present (i) the first analysis of the high-resolution ro-vibrational spectrum of aminoacetonitrile, where the three lowest excited states are observed and (ii) the search for vibrationally excited aminoacetonitrile with ALMA using the ReMoCA survey.
	
	This paper is structured as follows. Section~\ref{sec:exp} contains details about the experimental setup used to record the FIR spectrum.
	In Sect.~\ref{sec:spec}, we give an account of the spectroscopic properties of aminoacetonitrile, describe the spectral analysis and its results, and explain how new catalog entries were prepared. Section~\ref{sec:obs} reports our new astronomical observations of aminoacetonitrile. Finally, the results are discussed in Sect.~\ref{sec:discussion} and conclusions are drawn in Sect.~\ref{sec:concl}.

	%---------------------------------------------------------------
	
	\begin{figure*}[htb!]
		\centering
		\includegraphics[width=\textwidth]{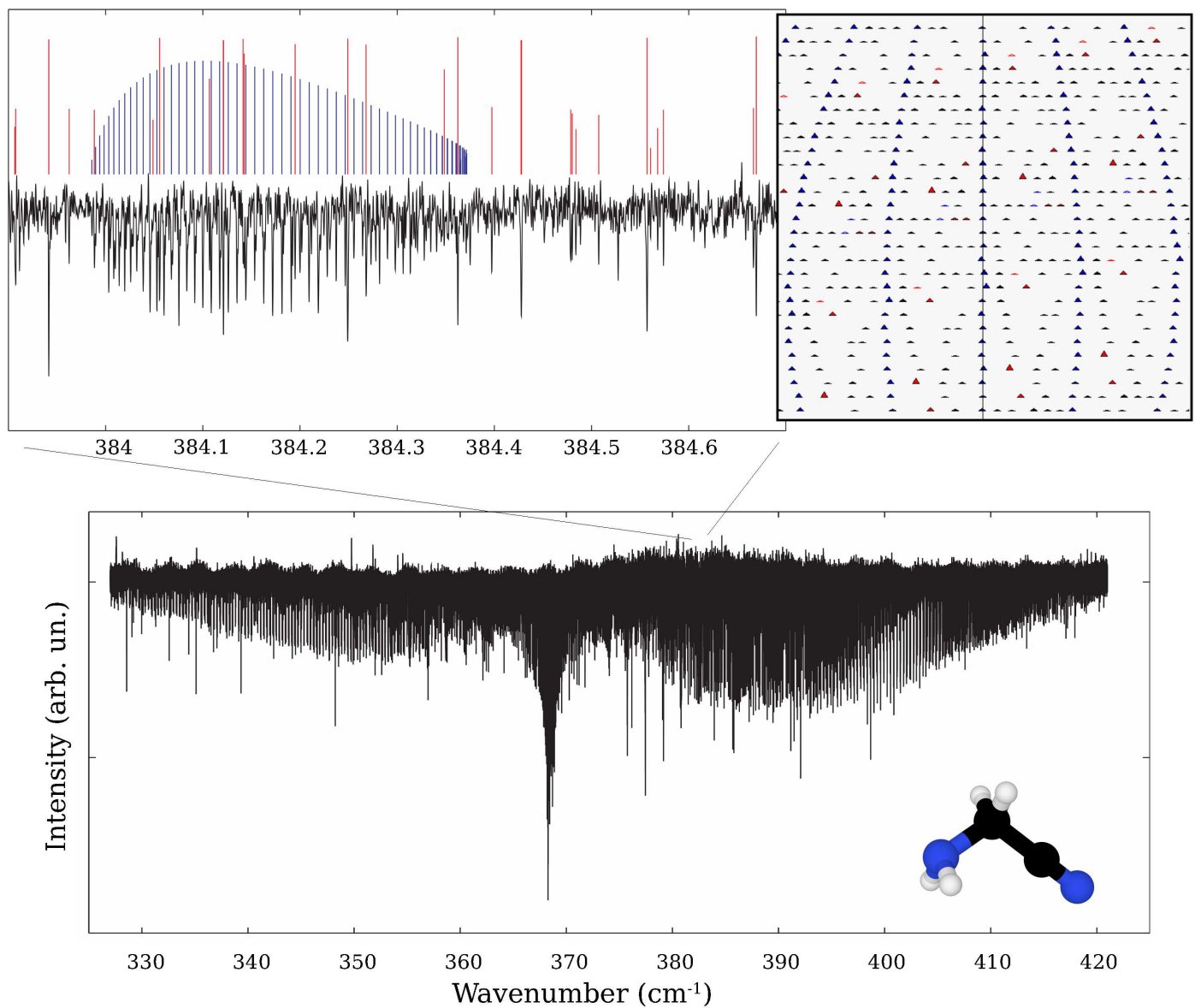}
		\caption{(\textit{Bottom panel}): overview of the $\nu_{17}$ band, centered around 368\,\wn, recorded at 3\,\textmu bar pressure. Residual water lines were blanked out from the spectrum. The molecular geometry of aminoacetonitrile is also shown with carbon in black, nitrogen in blue, and hydrogen in white. (\textit{Top left panel}): detail of the $^RQ_9$ branch. The black trace is the experimental spectrum, blue sticks represent the position and the intensity of the $^RQ_9$ components, while red sticks represent other $^RR$ transitions. (\textit{Top right panel}): Loomis-Wood diagram of the $^RQ_9$ branch using the same color legend. The wavenumber axis of the spectra ($x$) is plotted against consecutive $J$ transitions across the $y$-axis; each row of the plot is 0.02\,\wn-wide. The height of triangles is proportional to the spectral line intensity.}
		\label{fig:nu17}
	\end{figure*}

	\section{Experimental details}\label{sec:exp}
	
	The FIR spectrum of aminoacetonitrile was recorded between 100 and 500\,\wn on the AILES\footnote{\url{https://www.synchrotron-soleil.fr/en/beamlines/ailes}} beamline of the SOLEIL synchrotron facility.
	The bright synchrotron radiation was extracted and focused onto the entrance iris (2\,mm) of a Bruker IFS 125 Fourier transform (FT) interferometer equipped with a 6\,\textmu m Mylar-silicon composite beamsplitter and a liquid helium-cooled silicon bolometer \citep{Brubach2010}. 
	The interferometer was continuously evacuated to about 0.1\,\textmu bar to limit the absorption of atmospheric water. 
	Spectra were recorded in a White-type multipass cell whose optics were adjusted to attain a 150\,m optical path length \citep{Pirali2012} and isolated from the interferometer by 50\,\textmu m-thick polypropylene windows. 
	The sample of aminoacetonitrile ($\ge98$\,\% purity) was purchased from Sigma Aldrich and used without further purification.
	Two spectra were recorded with aminoacetonitrile at different pressures (1 and 3\,\textmu bar) and exploiting the highest resolution of the Bruker spectrometer (0.00102\,\wn).
	About 300 scans were co-added in order to improve the signal-to-noise ratio (S/N) of the spectra.
	
	The wavenumber-axis of the spectra has been calibrated using residual water absorption lines, whose reference wavenumbers were taken from \citet{MatsushimaCalibrationH2O} and \citet{Horneman2005}.
	Based on the dispersion after calibration, line frequencies are expected to be as accurate as 0.0001\,\wn.
	
	%---------------------------------------------------------------

	\section{Spectral analysis and results}\label{sec:spec}
	
	\subsection{Spectroscopic properties}
	
	Aminoacetonitrile is an asymmetric-top rotor close to the prolate limit ($\kappa=-0.965$).
	Its ro-vibrational energy levels can be modeled using a $S$-reduced Watson-type \citep{watson1977aspects} Hamiltonian:
	\begin{equation} \label{eq:ham}
	\mathscr{H} = \mathscr{H}_\mrm{rv} + \mathscr{H}_\mrm{cd} + \mathscr{H}_\mrm{Cor} \,,
	\end{equation}
	where $\mathscr{H}_\mrm{rv}$ is the ro-vibrational Hamiltonian containing the vibrational energy $E$ and the rotational constants $A$, $B$, and $C$ of a given vibrational state, while the $\mathscr{H}_\mrm{cd}$ term describes the centrifugal distortion effect during the molecular rotation. These two terms constitute the classic Hamiltonian for a semi-rigid rotor.
	The third term of the right side is the Coriolis Hamiltonian ($\mathscr{H}_\mrm{Cor}$) and accounts for the ro-vibrational interaction between states that are close in energy. The latter term has been introduced solely for the $\varv_{11}=1$ and $\varv_{18}=1$ states, which were found to perturb each other \citep[see][]{degli2017millimeter}.
	Conversely, the ro-vibrational energy levels of the ground and $\varv_{17}=1$ states could be reproduced at the experimental accuracy by using the standard semi-rigid Hamiltonian.
	
	\begin{table*}[htb!]
		\centering
		\caption{Summary of the vibrational modes studied in this work.}
		\label{tab:modes}
		\begin{tabular}{c ccccccc}
			\hline\hline \\[-1ex]
			Vibration  & Description & Energy\,\tablefootmark{a} & Symmetry & Envelope & Intensity\,\tablefootmark{b} & No. of lines & rms $\times 10^4$\,\tablefootmark{c} \\[0.5ex]
			&             & (\wn)  &          &          &         &                 & (\wn) \\[0.5ex]
			\hline \\[-1.5ex]
			$\nu_{11}$ & \ce{CCN} bending & 210.575841(5)     & $A^\prime$         & $a/b$ & w & 1110 & 1.3 \\[0.5ex]
			$\nu_{18}$ & \ce{NH2} torsion & 244.891525(3)     & $A^{\prime\prime}$ & $c$   & m/s & 6122 & 1.0 \\[0.5ex]
			$\nu_{17}$ & \ce{CH2}-\ce{NH2} torsion & 368.104656(3) & $A^{\prime\prime}$ & $c$  & m & 3704 & 1.0 \\[0.5ex]
			\hline\hline
		\end{tabular}
		\tablefoot{\tablefoottext{a}{Numbers in parenthesis represent the $1\sigma$ standard error of the constant in unit of the last digit.} \tablefoottext{b}{Abbreviations are used as follows: w = weak, m = medium, s = strong.} \tablefoottext{c}{Root-mean-square error from the final fit.} }
	\end{table*}
	
	The most stable conformer of aminoacetonitrile is the \textit{trans} form, i.e. with the \ce{NH2} group pointing towards the \ce{CCN} chain (as shown in the molecular structure of Fig.~\ref{fig:nu17}). This conformation has a $C_\mrm{s}$ symmetry; therefore, the vibrational modes of aminoacetonitrile are either symmetric ($A^\prime$) or antisymmetric ($A^{\prime\prime}$) with respect to the reflection in the molecular plane $ab$ formed by the four heavy atoms.
	Since the selection rules for ro-vibrational transitions are governed by the change of dipole moment components with the vibration, the appearance of the spectrum reflects the symmetry of vibrational modes.
	In particular, $A^\prime$ vibrations give rise to $a/b$-type hybrid bands, while $A^{\prime\prime}$ vibrational bands possess a $c$-type envelope.
	
	For a given vibrational state $\varv$, each rotational energy level is labeled as $J_{K_a,\,K_c}$.
	The selection rules for each type of ro-vibrational transition are:
	\begin{subequations} \label{eq:selrul}
		\begin{eqnarray} \label{eq:subsel}
		a\text{-type} \quad &\Delta K_a = 0,\,\pm 2,\,\ldots \,\,     &\Delta K_c = \pm 1,\,\pm 3,\,\ldots \,, \\
		b\text{-type} \quad &\Delta K_a = \pm 1,\,\pm 3,\,\ldots \,\, &\Delta K_c = \pm 1,\,\pm 3,\,\ldots \,, \\
		c\text{-type} \quad &\Delta K_a = \pm 1,\,\pm 3,\,\ldots \,\, &\Delta K_c = 0,\,\pm 2,\,\ldots \,.
		\end{eqnarray}
	\end{subequations}
	The allowed changes for the other quantum numbers are $\Delta\varv=1$ and $\Delta J= 0$ ($Q$), $-1$ ($P$), or $+1$ ($R$).
	
	\subsection{Assignment procedure}
	
	Initially, an input set of parameters constituted by the spectroscopic constants derived from rotational measurements \citep{degli2017millimeter} and vibrational energies from \citet{bak1975vibrational} was used to predict the FIR spectrum.
	On the ground that the $\varv_{17}=1$ state is isolated and unperturbed, the $\nu_{17}$ band was analyzed first.
	The bottom panel of Fig.~\ref{fig:nu17} shows an overview of the $\nu_{17}$ band as observed in the 3\,\textmu bar spectrum. As expected from its $A^{\prime\prime}$ symmetry, this band has a $c$-type contour characterized by a prominent $Q$ branch near its center.
	The assignment of some transitions belonging to the strong $^RQ_0$ branch\footnote{In this symbolism, the superscript denotes the change of $K_a$, the middle letter represents the change of $J$, while the subscript gives the $K_a$ value of the lower level.} allowed a first adjustment of the vibrational energy of the $\varv_{17}=1$ state, previously known with an uncertainty of a few wavenumbers.
	Once the vibrational energy had been refined, the spectral analysis was extended towards higher $J$ and $K_a$ transitions.
	The assignment procedure was performed with the Loomis-Wood for Windows (LWW) program package \citep{lodyga2007advanced}, a useful tool designed for the graphical analysis of ro-vibrational spectra.
	The main feature of the LWW package is the symbolic representation of sequences of transitions with Loomis-Wood diagrams; such plots provide a graphical representation of the spectrum where recognizable patterns belonging to the same branch appear in rows of a given width \citep{winnewisser1989interactive}.
	This procedure guarantees an easy and fast analysis, while internally checking the assignments by means of lower state combination differences (LSCD).
	With this method, several $P$, $Q$, and $R$ branches of the $\nu_{17}$ band have been assigned in the range 325--415\,\wn for a total number of $\sim$~3700 distinct lines. An excerpt of this band is shown in the upper panels of Fig.~\ref{fig:nu17}, together with its LW plot.
	
	Subsequently, we proceeded with the analysis of the $\nu_{18}$ band.
	This mode is of $A^{\prime\prime}$ symmetry too, thus producing a $c$-type contour analogous to that of the $\nu_{17}$ band.
	To avoid saturation issues, the strongest absorption features of this band were analyzed in the 1\,\textmu bar spectrum, while the 3\,\textmu bar spectrum was used to detect weaker transitions.
	Being the most intense band of the FIR region, more than 6\,000 ro-vibrational transitions could be assigned and added to our dataset.
	
	The analysis of the $\nu_{11}$ band, the lowest energetic mode around 210\,\wn, was more challenging under many aspects. Being of $A^\prime$ symmetry, this band must have an $a$-type and/or a $b$-type contour. While an \textit{a priori} evaluation is not always possible, both envelopes are less prominent than a $c$-type band in any case.
	Moreover, the $\nu_{11}$ band is expected to be 4-5 times weaker than the neighbour $\nu_{18}$ \citep{bernstein2004infrared}, whose spectral extent covers a huge range (195--305\,\wn).
	Nevertheless, the existence of a Coriolis resonance between these two states represents a great advantage, since it allows a determination of their vibrational energy difference. In their work, \citet{degli2017millimeter} derived a value of 34.3173(3)\,\wn.
	Combining it with the newly determined vibrational energy of the $\varv_{18}=1$ state, the $\nu_{11}$ band center could be estimated with sufficient accuracy, a mandatory requirement for the assignment of weak transitions in a crowded spectrum.
	About 35 $b$-type branches of transitions were identified in the 3\,\textmu bar spectrum, with a systematic blue shift from predictions of just 0.001\,\wn. Sequences of $a$-type transitions were searched for too, but no unequivocal evidence was found for any.
	Therefore, given the weaker nature of this band, only the strongest $b$-type transitions could be confidently assigned.
	In total, 1110 distinct lines of the $\nu_{11}$ band were included in the analysis.
	The main features of the three ro-vibrational bands are summarized in Table~\ref{tab:modes}.

	\subsection{Results from the analysis}
	
	\begin{table*}[htb!]
		\centering
		\caption{Spectroscopic constants of aminoacetonitrile determined for the ground, $\varv_{11}=1$, $\varv_{18}=1$, and $\varv_{17}=1$ states.}
		\label{tab:par}
		\begin{tabular}{lc cccc}
			\hline\hline \\[-1ex]
			Constant & Unit & GS & $\varv_{11}=1$ & $\varv_{18}=1$ & $\varv_{17}=1$ \\[0.5ex]
			\hline \\[-1.5ex]
			$   E$     & \wn  &     0.0          & 210.575842(5)   & 244.891525(3)  & 368.104657(3)  \\[0.5ex]
			$   A$     & MHz  &  30246.4909(9)   & 30301.42(8)     & 30344.49(8)    & 30143.688(2)   \\[0.5ex]
			$   B$     & MHz  &   4761.0626(1)   &  4776.4958(1)   &  4769.4229(1)  &  4764.2372(1)  \\[0.5ex]
			$   C$     & MHz  &   4310.7486(1)   &  4316.6461(1)   &  4314.4987(1)  &  4316.4332(1)  \\[0.5ex]
			$  D_J$    & kHz  &      3.0669(1)   &   3.04639(10)   &    3.07205(5)  &     3.0721(1)  \\[0.5ex]
			$ D_{JK}$  & kHz  &    -55.295(1)    &   -52.42(2)     &   -55.81(2)    &   -55.045(2)   \\[0.5ex]
			$  D_K$    & kHz  &    714.092(7)    &   714.092       &   713.5(1)     &   695.58(2)    \\[0.5ex]
			$  d_1$    & kHz  &     -0.67355(4)  &    -0.67477(5)  &    -0.67567(6) &    -0.67216(6) \\[0.5ex]
			$  d_2$    & kHz  &     -0.02993(1)  &    -0.034743(6) &    -0.02897(1) &    -0.02667(2) \\[0.5ex]
			$  H_J$    & mHz  &      9.56(3)     &     8.91(3)     &     9.31(1)    &     9.42(3)    \\[0.5ex]
			$ H_{JK}$  & Hz   &     -0.1249(4)   &    -0.1249      &    -0.1087(2)  &    -0.1227(7)  \\[0.5ex]
			$ H_{KJ}$  & Hz   &     -2.714(4)    &    -2.714       &   -2.714       &    -2.74(1)    \\[0.5ex]
			$  H_K$    & Hz   &     53.27(2)     &    53.27        &   53.27        &    47.65(4)    \\[0.5ex]
			$  h_1$    & mHz  &      3.88(1)     &     3.66(1)     &    3.60(2)     &     3.78(2)    \\[0.5ex]
			$  h_2$    & mHz  &      0.476(6)    &     0.476       &    0.476       &     0.436(9)   \\[0.5ex]
			$  h_3$    & mHz  &      0.0503(8)   &     0.0503      &    0.0503      &     0.0503     \\[0.5ex]
			$  L_J$    & \textmu Hz & -0.037(3)  &    \ldots       &    \ldots      &   \ldots       \\[0.5ex]
			$ L_{JJK}$ & \textmu Hz &  0.47(6)   &    \ldots       &    \ldots      &   \ldots       \\[0.5ex]
			$ L_{KKJ}$ & mHz  &        0.167(8)  &    \ldots       &    \ldots      &   \ldots       \\[0.5ex]
			$  L_K$    & mHz  &       -4.43(2)   &    \ldots       &    \ldots      &   \ldots       \\[0.5ex]
			$  l_1$    & \textmu Hz & -0.021(1)  &    \ldots       &    \ldots      &   \ldots       \\[0.5ex]
			$  l_2$    & \textmu Hz & -0.0054(9) &    \ldots       &    \ldots      &   \ldots       \\[0.5ex]
			\hline \\[-1.5ex]
			$   G_a   $  & MHz &       \ldots    &     \mcl{2}{c}{17070.(2)   }  &       \ldots     \\[0.5ex]
			$  G_a^J$    & kHz &       \ldots    &     \mcl{2}{c}{  25.9(5)   }  &       \ldots     \\[0.5ex]
			$ G_a^{JK}$  &  Hz &       \ldots    &     \mcl{2}{c}{   -31.2(3) }  &       \ldots     \\[0.5ex]
			$ G_a^{K}$   & MHz &       \ldots    &     \mcl{2}{c}{  -1.916(3) }  &       \ldots     \\[0.5ex]
			$ G_a^{KK}$  & kHz &       \ldots    &     \mcl{2}{c}{   0.214(3) }  &       \ldots     \\[0.5ex]
			$ G_a^{JKK}$ &  Hz &       \ldots    &     \mcl{2}{c}{   0.0466(4)}  &       \ldots     \\[0.5ex]
			$ G_a^{KKK}$ &  Hz &       \ldots    &     \mcl{2}{c}{  -0.195(2) }  &       \ldots     \\[0.5ex]
			\hline\hline
		\end{tabular}
		\tablefoot{Numbers in parenthesis represent the $1\sigma$ standard error of the constant in unit of the last digit. Constants without error are held fixed to the corresponding ground state value.}
	\end{table*}
		
	The newly observed ro-vibrational transitions were analyzed in a weighted least-squares procedure performed with the \texttt{SPFIT} subprogram of the \texttt{CALPGM} suite \citep{pickett1991}.
	Pure rotational transitions of the ground state \citep{macdonald1972microwave,pickett1973microwave,brown1977quadrupole,bogey1990millemeter,motoki2013submillimeter} and excited states of aminoacetonitrile \citep{degli2017millimeter} were also included in the analysis.
	In the least-squares procedure, each datum was weighted proportionally to the inverse square of its uncertainty. Literature data were used with the errors quoted in the original papers, while our new data were mostly given uncertainties of $1\times10^{-4}$\,\wn.
	A conservative uncertainty of $1.5\times10^{-4}$\,\wn was given to the transition frequencies of the weaker $\nu_{11}$ band, to account for frequent overlaps with other lines.
	
	The dataset includes more than 2\,000 pure rotational transitions and about 11\,000 ro-vibrational lines, probing energy levels with $J$ values up to 80 and $K_a$ up to 25.
	The overall standard deviation of the fit ($\sigma$) is 0.99, meaning that our model satisfactorily reproduces the observed transition frequencies at their experimental accuracy. The root-mean-square (rms) error of FIR lines is $1.1\times10^{-4}$\,\wn, whereas rotational data show an rms error of 40\,kHz.
	The set of spectroscopic parameters derived in the present analysis is collected in Table~\ref{tab:par}.

	\subsection{Partition function}
	
	The parameters of Table~\ref{tab:par} have been input in the \texttt{SPCAT} subroutine \citep{pickett1991} to compute the ro-vibrational partition function of aminoacetonitrile analytically.
	The temperature-dependence of the partition function $Q$ has been calculated for three different energy level manifolds, one including the rotational levels of the ground state only and the other two accounting for the energy levels of all vibrational states below 400 and 1600\,\wn, respectively.
	As expected, the values obtained for the ground state of aminoacetonitrile are almost identical to those reported by \citet{motoki2013submillimeter} and in the Cologne database for molecular spectroscopy (CDMS, \citealt{endres2016cologne}).
	
	On the other hand, the manifold containing all states below 400\,\wn (referred to as ``Manifold 1'' in Table~\ref{tab:qvibrot}) gives quite different results from Table~4 of \citet{degli2017millimeter}. However, the discrepancy is most likely due to the fact that \citet{degli2017millimeter} evaluated the contribution of the $\varv_{11}=1$ and $\varv_{18}=1$ states considering only their energy difference, but omitting their absolute vibrational energies \footnote{This led to an overestimation of $Q$, as it implied that the vibrational energies of $\varv_{11}=1$ and $\varv_{18}=1$ were 0 and 34.3173(3)\,\wn, respectively.}.
	We regard our calculations to be robust, as they are obtained without any assumptions.
	
	To compute the ro-vibrational partition function of the manifold containing all states up to 1600\,\wn (referred to as ``Manifold 2'' in Table~\ref{tab:qvibrot}), we have applied a vibrational correction to our new ground state partition function values. The correction factor was calculated at each temperature summing up the contribution of all the vibrational states below 1600\,\wn. Since we have no direct information about the vibrational energies of the states between 400 and 1600\,\wn, they were either estimated under the harmonic approximation or taken from low resolution measurements \citep{bak1975vibrational}. Higher-energy levels were not considered in the calculation because they do not contribute significantly to $Q$ even at 300\,K.
	The vibration-rotation partition functions of the three manifolds of aminoacetonitrile, computed at temperatures between 2.725 and 300\,K, are listed in Table~\ref{tab:qvibrot}.
	
	Based on our results, we have prepared new catalog entries for aminoacetonitrile in the ground and excited states.
	The line intensity of each transition was calculated using the partition function values from Col.~4 of Table~\ref{tab:qvibrot}, while the permanent dipole moment values $\mu_a=2.577(7)$\,D and $\mu_b=0.575(1)$\,D were taken from \citet{pickett1973microwave} and assumed to not change upon the vibrational states.
	
	\begin{table}[htb!]
		\centering
		\caption{Ro-vibrational partition function values computed at different temperatures.}
		\label{tab:qvibrot}
		\begin{tabular}{c ccc}
			\hline\hline \\[-1ex]
			Temperature (K) & GS only & Manifold 1 \tablefootmark{a} & Manifold 2 \tablefootmark{b} \\[0.5ex]
			\hline \\[-1.5ex]
			300.000 &  35201.5484  &  64864.3838  & 107688.6464 \\[0.5ex]
			225.000 &  22898.2410  &  35787.1778  & 45029.6653  \\[0.5ex]
			150.000 &  12460.9459  &  15662.3573  & 16455.8252  \\[0.5ex]
			75.000 &   4403.3095  &   4524.5423  &  4527.3991  \\[0.5ex]
			37.500 &   1557.4422  &   1558.0551  &  1558.0551  \\[0.5ex]
			18.750 &    551.5089  &    551.5090  &   551.5090  \\[0.5ex]
			9.375 &    195.6798  &    195.6798  &   195.6798  \\[0.5ex]
			5.000 &     76.7031  &     76.7031  &    76.7031  \\[0.5ex]
			2.725 &     31.2184  &     31.2184  &    31.2184  \\[0.5ex]
			\hline\hline
		\end{tabular}
		\tablefoot{\tablefoottext{a}{Includes all states up to 400\,\wn; namely the ground, $\varv_{11}=1$, $\varv_{18}=1$, and $\varv_{17}=1$ states.} \tablefoottext{b}{Includes all states up to 1600\,\wn.}}
	\end{table}
	
	%---------------------------------------------------------------
	\section{Astronomical observations}\label{sec:obs}
	
	\subsection{Observations}
	\label{ss:observations}
	
	We used the ReMoCA imaging spectral line survey carried out toward the protostellar cluster Sgr~B2(N) with ALMA. Details about the observational setup and data reduction of this survey were reported in \citet{Belloche19}. In short, the observations covered the frequency range from 84.1~GHz to 114.4~GHz with five tunings, with a spectral resolution of 488~kHz (1.7 to 1.3~km~s$^{-1}$). The angular resolution (HPBW) varied between $\sim$0.3$\arcsec$ and $\sim$0.8$\arcsec$, with a median value of 0.6$\arcsec$, which corresponds to $\sim$4900~au at the distance of Sgr~B2. The rms sensitivity ranged from 0.35~mJy~beam$^{-1}$ to 1.1~mJy~beam$^{-1}$, with a median value of 0.8~mJy~beam$^{-1}$. The field was centered at ($\alpha, \delta$)$_{\rm J2000}$= ($17^{\rm h}47^{\rm m}19{\fs}87, -28^\circ22'16{\farcs}0$), a position that is half-way between the two main hot molecular cores, Sgr~B2(N1) and Sgr~B2(N2) which are separated by 4.9$\arcsec$ or $\sim$0.2~pc. Here we analyze the spectrum at the offset position Sgr~B2(N1S) defined by \citet{Belloche19} to reduce the optical depth of the continuum emission, which is partially optically thick toward the peak position of the main hot core Sgr~B2(N1). Sgr~B2(N1S) is located at ($\alpha, \delta$)$_{\rm J2000}$= ($17^{\rm h}47^{\rm m}19{\fs}870$, $-28^\circ22\arcmin19{\farcs}48$), about 1$\arcsec$ to the south of Sgr~B2(N1).
	
	In \citet{Belloche19}, the continuum and line emission of the ReMoCA survey was separated in the image plane because the hot cores detected in the field of view have different systemic velocities and some of the spectra are close to the confusion limit, the combination of both properties implying that splitting the continuum and line emission in the uv plane is fundamentally impossible because in basically every spectral channel there is somewhere in the field of view some line emission. The continuum and line emission separation performed in \citet{Belloche19} was still preliminary. We made some progress with our algorithm that performs this splitting across the whole field of view and we use the newly obtained spectra for the present analysis. There are still some limitations due to the quality of the spectral baselines which we hope to improve in the future.
	
	We modeled the spectrum of Sgr~B2(N1S) with the software Weeds \citep[][]{Maret11} under the assumption of local thermodynamic equilibrium (LTE), which is appropriate given the high densities that characterize the regions where hot-core emission is detected in Sgr~B2(N) \citep[$>1 \times 10^{7}$~cm$^{-3}$, see][]{Bonfand19}.
	We derived a best-fit synthetic spectrum of each molecule separately, and then added the contributions of all identified molecules together. Each species is modeled with a set of five parameters: size of the emitting region ($\theta_{\rm s}$), column density ($N$), temperature ($T_{\rm rot}$), linewidth ($\Delta V$), and velocity offset ($V_{\rm off}$) with respect to the assumed systemic velocity of the source ($V_{\rm sys}=62$~km~s$^{-1}$).
	
	\subsection{Detection of vibrationally excited aminoacetonitrile}
	\label{ss:detection}
	
	\begin{figure*}[!ht]
		\centerline{\resizebox{0.82\hsize}{!}{\includegraphics[angle=0]{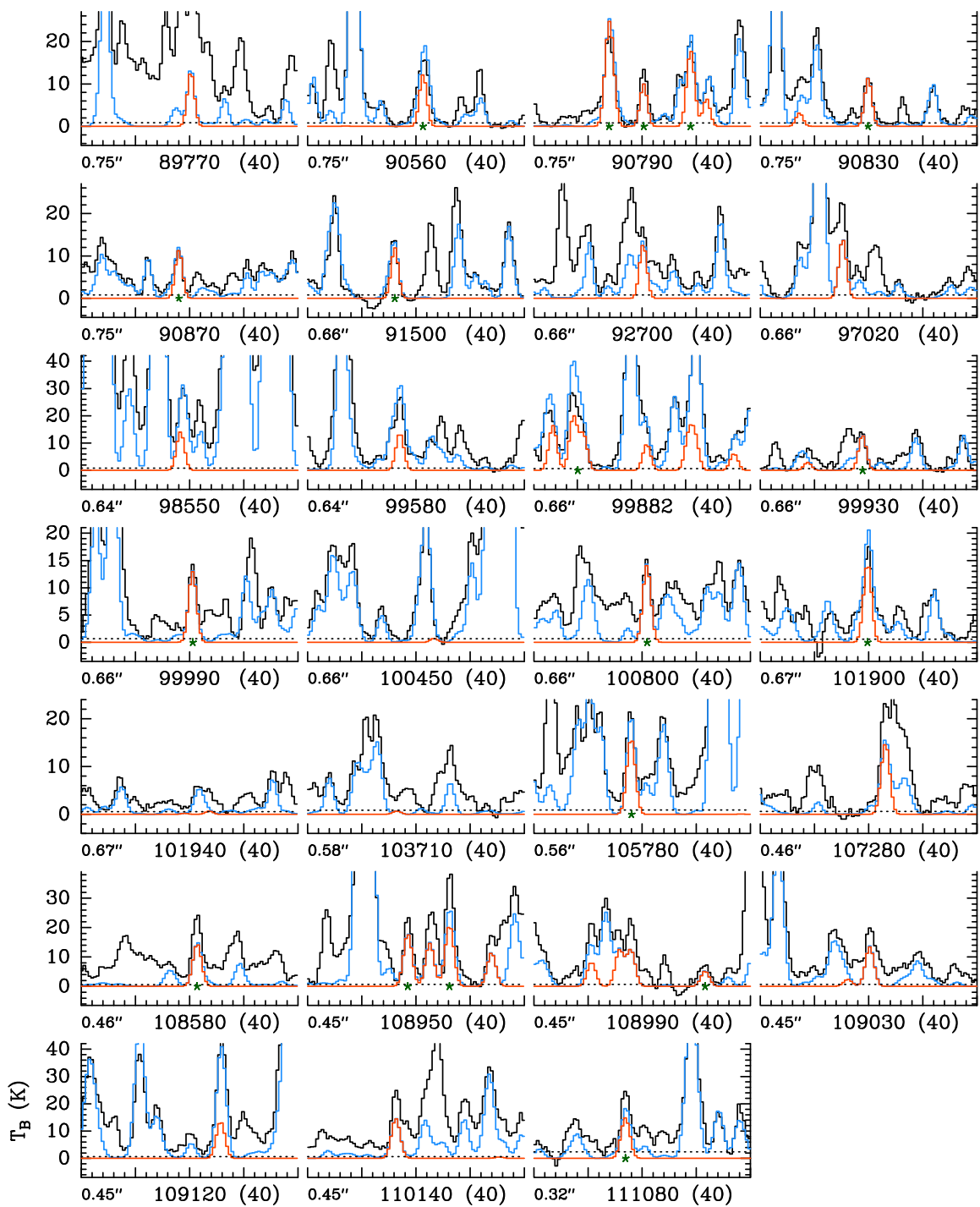}}}
		\caption{Transitions of \ce{NH2CH2CN}, $\varv = 0$ covered by our ALMA survey. The best-fit LTE synthetic spectrum of \ce{NH2CH2CN}, $\varv = 0$ is displayed in red and overlaid on the observed spectrum of Sgr~B2(N1S) shown in black. The blue synthetic spectrum contains the contributions of all molecules identified in our survey so far, including the species shown in red. The central frequency and width are indicated in MHz below each panel. The angular resolution (HPBW) is also indicated. The $y$-axis is labeled in brightness temperature units (K). The dotted line indicates the $3\sigma$ noise level. The green stars mark the aminoacetonitrile lines that we consider as unambiguously detected because they suffer from little contamination by emission from other species.}
		\label{f:spec_aan_ve0}
	\end{figure*}
	
	\begin{table*}
		\begin{center}
			\caption{
				Spectroscopic parameters and integrated intensities of aminoacetonitrile transitions detected toward Sgr~B2(N1S) in the ReMoCA survey.
			}
			\label{t:aan_det_int}
			\vspace*{0.0ex}
			\begin{tabular}{lcrcccccrr}
				\hline\hline
				\multicolumn{1}{c}{State} & \multicolumn{1}{c}{Transition} & \multicolumn{1}{c}{Frequency} & \multicolumn{1}{c}{$\Delta f$\tablefootmark{a}}  & \multicolumn{1}{c}{$A_{\rm ul}$\tablefootmark{b}} & \multicolumn{1}{c}{$E_{\rm u}$\tablefootmark{c}}  & \multicolumn{1}{c}{$g_{\rm u}$\tablefootmark{d}} & \multicolumn{1}{c}{$I_{\rm obs}$\tablefootmark{e}} & \multicolumn{1}{c}{$I_{\rm mod}$\tablefootmark{f}} & \multicolumn{1}{c}{$I_{\rm all}$\tablefootmark{g}} \\ 
				& \multicolumn{1}{c}{$J_{K_a,K_c}$} & \multicolumn{1}{c}{\small (MHz)} & \multicolumn{1}{c}{\small (kHz)} & \multicolumn{1}{c}{\small (10$^{-5}$ s$^{-1}$)} & \multicolumn{1}{c}{\small (K)} & & \multicolumn{1}{c}{\small (K km s$^{-1}$)} & \multicolumn{2}{c}{\small (K km s$^{-1}$)} \\ 
				\hline
				$\varv=0$ & 10$_{2,9}$ -- 9$_{2,8}$ & 90561.326 &   2 &  2.62 &    28.9 & 21 & 127.9(12) &  67.2 & 137.9 \\ 
				$\varv=0$ & 10$_{6,4}$ -- 9$_{6,3}$ & 90783.532 &   2 &  1.76 &    68.3 & 21 & 169.7(13) & 161.3 & 186.4 \\ 
				$\varv=0$ & 10$_{6,5}$ -- 9$_{6,4}$ & 90783.532 &   2 &  1.76 &    68.3 & 21 & -- & -- & -- \\ 
				$\varv=0$ & 10$_{5,6}$ -- 9$_{5,5}$ & 90784.276 &   2 &  2.07 &    54.8 & 21 & -- & -- & -- \\ 
				$\varv=0$ & 10$_{5,5}$ -- 9$_{5,4}$ & 90784.280 &   2 &  2.07 &    54.8 & 21 & -- & -- & -- \\ 
				$\varv=0$ & 10$_{7,3}$ -- 9$_{7,2}$ & 90790.252 &   2 &  1.41 &    84.3 & 21 &  81.3(12) &  54.6 &  56.0 \\ 
				$\varv=0$ & 10$_{7,4}$ -- 9$_{7,3}$ & 90790.252 &   2 &  1.41 &    84.3 & 21 & -- & -- & -- \\ 
				$\varv=0$ & 10$_{4,7}$ -- 9$_{4,6}$ & 90798.680 &   2 &  2.31 &    43.7 & 21 & 159.0(12) & 105.6 & 163.5 \\ 
				$\varv=0$ & 10$_{4,6}$ -- 9$_{4,5}$ & 90799.244 &   2 &  2.31 &    43.7 & 21 & -- & -- & -- \\ 
				$\varv=0$ & 10$_{3,8}$ -- 9$_{3,7}$ & 90829.941 &   2 &  2.51 &    35.1 & 21 &  76.2(12) &  62.1 &  63.5 \\ 
				$\varv=0$ & 10$_{3,7}$ -- 9$_{3,6}$ & 90868.035 &   2 &  2.51 &    35.1 & 21 &  73.6(12) &  62.1 &  75.3 \\ 
				$\varv=0$ & 10$_{2,8}$ -- 9$_{2,7}$ & 91496.111 &   2 &  2.71 &    29.0 & 21 &  79.4(12) &  65.5 &  87.1 \\ 
				$\varv=0$ & 11$_{5,7}$ -- 10$_{5,6}$ & 99869.300 &   2 &  2.92 &    59.6 & 23 & 253.4(11) & 177.1 & 349.7 \\ 
				$\varv=0$ & 11$_{5,6}$ -- 10$_{5,5}$ & 99869.310 &   2 &  2.92 &    59.6 & 23 & -- & -- & -- \\ 
				$\varv=0$ & 11$_{7,4}$ -- 10$_{7,3}$ & 99871.145 &   2 &  2.19 &    89.1 & 23 & -- & -- & -- \\ 
				$\varv=0$ & 11$_{7,5}$ -- 10$_{7,4}$ & 99871.145 &   2 &  2.19 &    89.1 & 23 & -- & -- & -- \\ 
				$\varv=0$ & 11$_{3,9}$ -- 10$_{3,8}$ & 99928.882 &   2 &  3.41 &    39.9 & 23 &  88.1(9) &  70.5 &  76.3 \\ 
				$\varv=0$ & 11$_{3,8}$ -- 10$_{3,7}$ & 99990.564 &   2 &  3.42 &    39.9 & 23 &  82.1(9) &  70.7 &  74.1 \\ 
				$\varv=0$ & 11$_{2,9}$ -- 10$_{2,8}$ & 100800.880 &   2 &  3.66 &    33.8 & 23 &  91.6(9) &  76.5 &  79.9 \\ 
				$\varv=0$ & 11$_{1,10}$ -- 10$_{1,9}$ & 101899.798 &   2 &  3.88 &    30.6 & 23 & 117.7(8) &  79.0 & 139.0 \\ 
				$\varv=0$ & 12$_{1,12}$ -- 11$_{1,11}$ & 105777.967 &   3 &  4.36 &    34.3 & 25 & 133.3(12) &  87.4 & 116.8 \\ 
				$\varv=0$ & 12$_{2,11}$ -- 11$_{2,10}$ & 108581.403 &   2 &  4.62 &    38.9 & 25 & 156.5(8) &  78.5 &  86.7 \\ 
				$\varv=0$ & 12$_{6,7}$ -- 11$_{6,6}$ & 108948.518 &   3 &  3.60 &    78.4 & 25 & 135.7(9) & 100.3 & 105.0 \\ 
				$\varv=0$ & 12$_{6,6}$ -- 11$_{6,5}$ & 108948.518 &   3 &  3.60 &    78.4 & 25 & -- & -- & -- \\ 
				$\varv=0$ & 12$_{5,8}$ -- 11$_{5,7}$ & 108956.201 &   2 &  3.97 &    64.8 & 25 & 204.4(8) & 112.0 & 148.0 \\ 
				$\varv=0$ & 12$_{5,7}$ -- 11$_{5,6}$ & 108956.224 &   2 &  3.97 &    64.8 & 25 & -- & -- & -- \\ 
				$\varv=0$ & 12$_{10,2}$ -- 11$_{10,1}$ & 109001.594 &   3 &  1.47 &   157.1 & 25 &  49.7(8) &  28.2 &  29.1 \\ 
				$\varv=0$ & 12$_{10,3}$ -- 11$_{10,2}$ & 109001.594 &   3 &  1.47 &   157.1 & 25 & -- & -- & -- \\ 
				$\varv=0$ & 12$_{1,11}$ -- 11$_{1,10}$ & 111076.902 &   2 &  5.05 &    36.0 & 25 & 154.7(30) &  81.1 & 111.4 \\ 
				\hline 
				$\varv_{11}=1$ & 10$_{7,3}$ -- 9$_{7,2}$ & 91000.390 &   2 &  1.38 &   386.8 & 21 &  13.5(9) &  10.4 &  13.1 \\ 
				$\varv_{11}=1$ & 10$_{7,4}$ -- 9$_{7,3}$ & 91000.390 &   2 &  1.38 &   386.8 & 21 & -- & -- & -- \\ 
				$\varv_{11}=1$ & 11$_{7,4}$ -- 10$_{7,3}$ & 100102.529 &   2 &  2.15 &   391.6 & 23 &  60.7(11) &  39.9 &  48.3 \\ 
				$\varv_{11}=1$ & 11$_{7,5}$ -- 10$_{7,4}$ & 100102.529 &   2 &  2.15 &   391.6 & 23 & -- & -- & -- \\ 
				$\varv_{11}=1$ & 11$_{5,7}$ -- 10$_{5,6}$ & 100104.591 &   2 &  2.90 &   362.3 & 23 & -- & -- & -- \\ 
				$\varv_{11}=1$ & 11$_{5,6}$ -- 10$_{5,5}$ & 100104.602 &   2 &  2.90 &   362.3 & 23 & -- & -- & -- \\ 
				$\varv_{11}=1$ & 11$_{1,10}$ -- 10$_{1,9}$ & 102166.422 &   2 &  3.91 &   333.7 & 23 &  37.1(7) &  17.7 &  26.5 \\ 
				$\varv_{11}=1$ & 12$_{5,8}$ -- 11$_{5,7}$ & 109213.457 &   2 &  3.94 &   367.6 & 25 &  43.5(8) &  38.8 &  51.1 \\ 
				$\varv_{11}=1$ & 12$_{5,7}$ -- 11$_{5,6}$ & 109213.483 &   2 &  3.94 &   367.6 & 25 & -- & -- & -- \\ 
				$\varv_{11}=1$ & 12$_{8,4}$ -- 11$_{8,3}$ & 109214.554 &   2 &  2.60 &   415.2 & 25 & -- & -- & -- \\ 
				$\varv_{11}=1$ & 12$_{8,5}$ -- 11$_{8,4}$ & 109214.554 &   2 &  2.60 &   415.2 & 25 & -- & -- & -- \\ 
				\hline 
				$\varv_{18}=1$ & 10$_{6,5}$ -- 9$_{6,4}$ & 90906.708 &   2 &  1.74 &   421.4 & 21 &  48.4(11) &  28.0 &  32.6 \\ 
				$\varv_{18}=1$ & 10$_{6,4}$ -- 9$_{6,3}$ & 90906.708 &   2 &  1.74 &   421.4 & 21 & -- & -- & -- \\ 
				$\varv_{18}=1$ & 10$_{5,6}$ -- 9$_{5,5}$ & 90906.982 &   1 &  2.05 &   407.6 & 21 & -- & -- & -- \\ 
				$\varv_{18}=1$ & 10$_{5,5}$ -- 9$_{5,4}$ & 90906.986 &   1 &  2.05 &   407.6 & 21 & -- & -- & -- \\ 
				\hline
			\end{tabular}
		\end{center}
		\vspace*{-2.5ex}
		\tablefoot{
			\tablefoottext{a}{Frequency uncertainty.}
			\tablefoottext{b}{Einstein coefficient for spontaneous emission.}
			\tablefoottext{c}{Upper-level energy.}
			\tablefoottext{d}{Upper-level degeneracy.}
			\tablefoottext{e}{Integrated intensity of the observed spectrum in brightness temperature scale. The statistical standard deviation is given in parentheses in unit of the last digit.}
			\tablefoottext{f}{Integrated intensity of the synthetic spectrum of NH$_2$CH$_2$CN.}
			\tablefoottext{g}{Integrated intensity of the model that contains the contribution of all identified molecules, including NH$_2$CH$_2$CN. In the last three columns, a value followed by dashes in the following lines represents the intensity integrated over a group of transitions that are not resolved in the astronomical spectrum.}
		}
	\end{table*}

\begin{figure*}[!ht]
		\centerline{\resizebox{0.82\hsize}{!}{\includegraphics[angle=0]{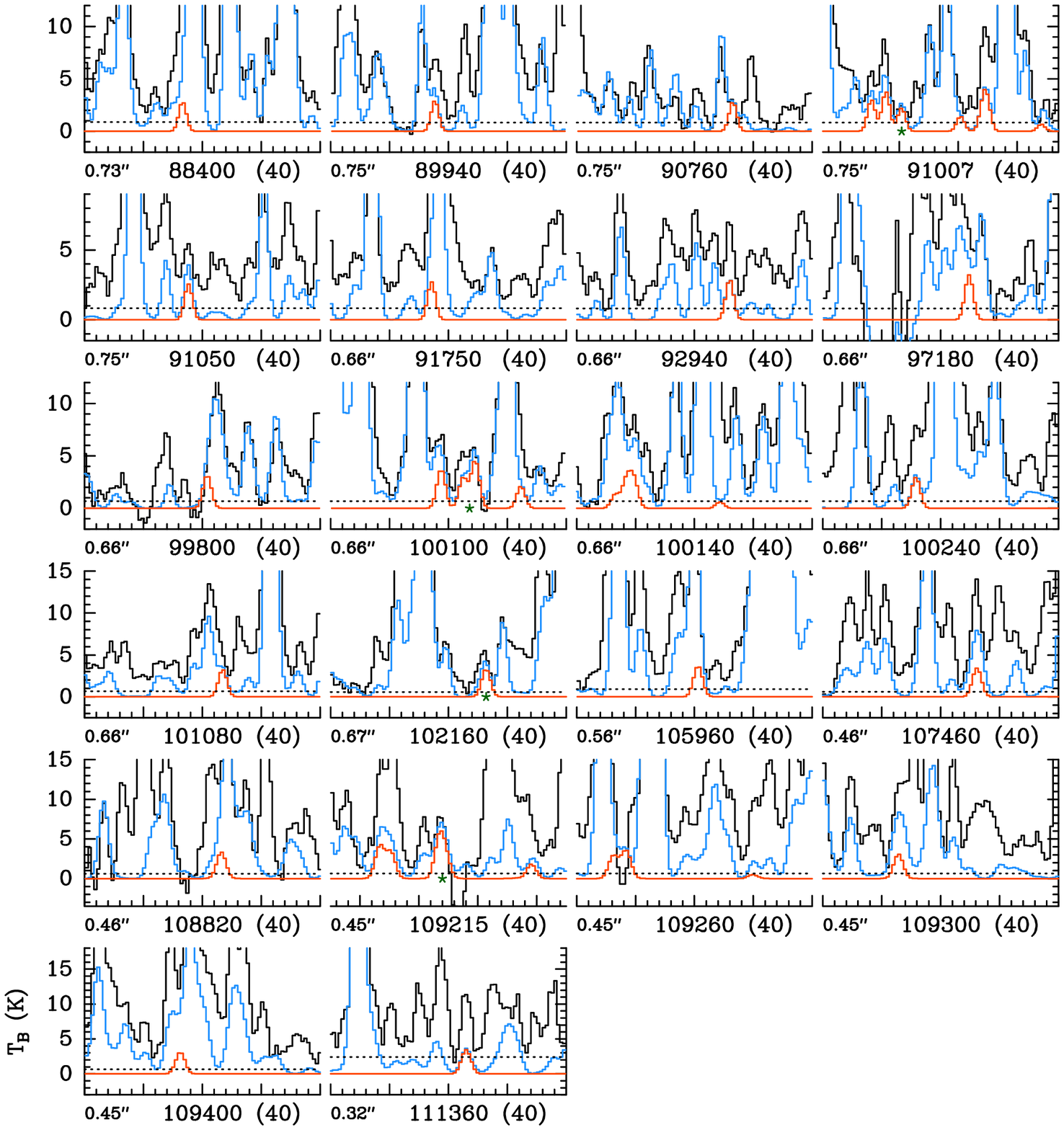}}}
		\caption{Same as Fig.~\ref{f:spec_aan_ve0} but for \ce{NH2CH2CN}, 
			$\varv_{11}=1$.}
		\label{f:spec_aan_v11e1}
	\end{figure*}

\begin{figure*}[!ht]
		\centerline{\resizebox{0.82\hsize}{!}{\includegraphics[angle=0]{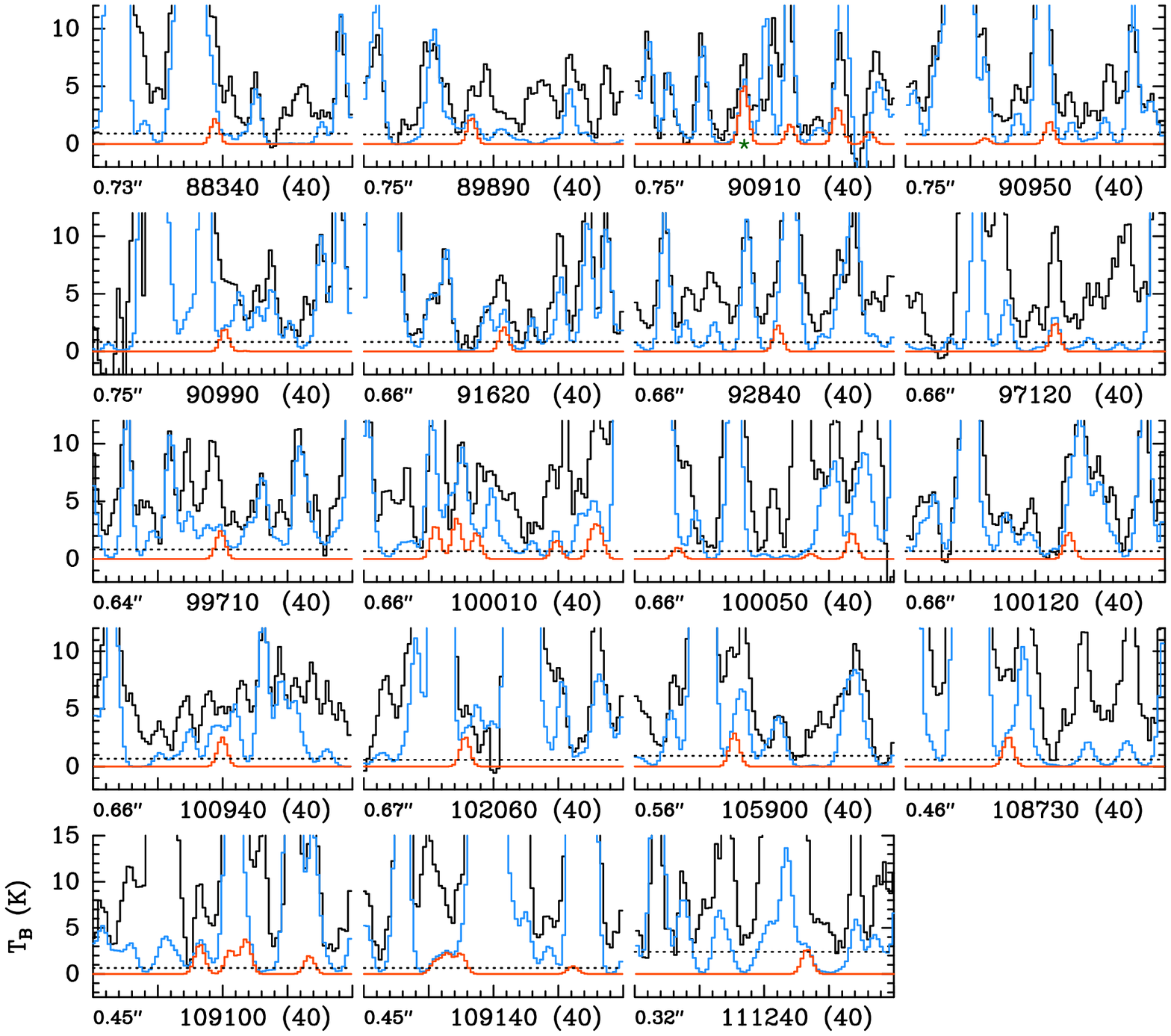}}}
		\caption{Same as Fig.~\ref{f:spec_aan_ve0} but for \ce{NH2CH2CN}, 
			$\varv_{18}=1$.}
		\label{f:spec_aan_v18e1}
	\end{figure*}

	We used the spectroscopic information derived in Sect.~\ref{sec:spec} to search for rotational transitions from within vibrationally excited states of aminoacetonitrile in the ReMoCA spectrum of Sgr~B2(N1S). As a first step, we identified a dozen of transitions of aminoacetonitrile in its vibrational ground state that suffer from little contamination by emission of other species. These detected transitions are marked with a star in Fig.~\ref{f:spec_aan_ve0} and are listed in Table~\ref{t:aan_det_int}. The best-fit LTE synthetic spectrum that we obtained for aminoacetonitrile is shown in red in Fig.~\ref{f:spec_aan_ve0} and its parameters are reported in Table~\ref{t:coldens}. Using the same parameters, we could identify four and one rotational transitions from within the vibrationally excited states $\varv_{11}=1$ and $\varv_{18}=1$, respectively. These transitions are marked with a star in Figs.~\ref{f:spec_aan_v11e1} and \ref{f:spec_aan_v18e1}, respectively, and are listed in Table~\ref{t:aan_det_int}.
	
	\begin{table*}[!ht]
		\begin{center}
			\caption{
				Parameters of our best-fit LTE model of aminoacetonitrile toward Sgr~B2(N1S).
			}
			\label{t:coldens}
			\vspace*{-1.2ex}
			\begin{tabular}{lcrcccccc}
				\hline\hline
				\multicolumn{1}{c}{Vib.} & \multicolumn{1}{c}{Status\tablefootmark{a}} & \multicolumn{1}{c}{$N_{\rm det}$\tablefootmark{b}} & \multicolumn{1}{c}{$\theta_{\rm s}$\tablefootmark{c}} & \multicolumn{1}{c}{$T_{\mathrm{rot}}$\tablefootmark{d}} & \multicolumn{1}{c}{$N$\tablefootmark{e}} & \multicolumn{1}{c}{$F_{\rm vib}$\tablefootmark{f}} & \multicolumn{1}{c}{$\Delta V$\tablefootmark{g}} & \multicolumn{1}{c}{$V_{\mathrm{off}}$\tablefootmark{h}} \\ 
				\multicolumn{1}{c}{state} & & & \multicolumn{1}{c}{\small ($''$)} & \multicolumn{1}{c}{\small (K)} & \multicolumn{1}{c}{\small (cm$^{-2}$)} & & \multicolumn{1}{c}{\small (km~s$^{-1}$)} & \multicolumn{1}{c}{\small (km~s$^{-1}$)} \\ 
				\hline
				$\varv=0$ & d & 18 &  2.0 &  200 &  1.1 (17) & 1.00 & 5.0 & 0.0 \\ 
				$\varv_{11}=1$ & d & 4 &  2.0 &  200 &  1.1 (17) & 1.00 & 5.0 & 0.0 \\ 
				$\varv_{18}=1$ & d & 1 &  2.0 &  200 &  1.1 (17) & 1.00 & 5.0 & 0.0 \\ 
				$\varv_{17}=1$ & n & 0 &  2.0 &  200 &  1.1 (17) & 1.00 & 5.0 & 0.0 \\ 
				$\varv_{11}=2$ & n & 0 &  2.0 &  200 &  1.1 (17) & 1.00 & 5.0 & 0.0 \\ 
				\hline 
			\end{tabular}
		\end{center}
		\vspace*{-2.5ex}
		\tablefoot{
			\tablefoottext{a}{d: detection, n: nondetection.}
			\tablefoottext{b}{Number of detected lines \citep[conservative estimate, see Sect.~3 of][]{Belloche16}. One line of a given species may mean a group of transitions of that species that are blended together.}
			\tablefoottext{c}{Source diameter (\textit{FWHM}).}
			\tablefoottext{d}{Rotational temperature.}
			\tablefoottext{e}{Total column density of the molecule. $x$ ($y$) means $x \times 10^y$. An identical value for all listed vibrational states means that LTE is an adequate description of the vibrational excitation.}
			\tablefoottext{f}{Correction factor that was applied to the column density to account for the contribution of vibrationally excited states, in the cases where this contribution was not included in the partition function of the spectroscopic predictions.}
			\tablefoottext{g}{Linewidth (\textit{FWHM}).}
			\tablefoottext{h}{Velocity offset with respect to the assumed systemic velocity of Sgr~B2(N1S), $V_{\mathrm{sys}} = 62$ km~s$^{-1}$.}
		}
	\end{table*}

	The model assumes a diameter (FWHM) of 2$\arcsec$ for the emission of aminoacetonitrile, as was done for other species reported by \citet{Belloche19} toward Sgr~B2(N1S). This size is approximately three times bigger than the angular resolution of the ReMoCA survey and thus its exact value does not matter much for the derivation of the column density. A two-dimensional Gaussian fit to the integrated intensity maps of the six least contaminated transitions of aminoacetonitrile yields a mean size of $\sim$1.9$\arcsec$ with an rms dispersion of $\sim$0.15$\arcsec$, consistent with our size assumption. 
	
\begin{figure}[!ht]
		\centerline{\resizebox{0.95\hsize}{!}{\includegraphics[angle=0]{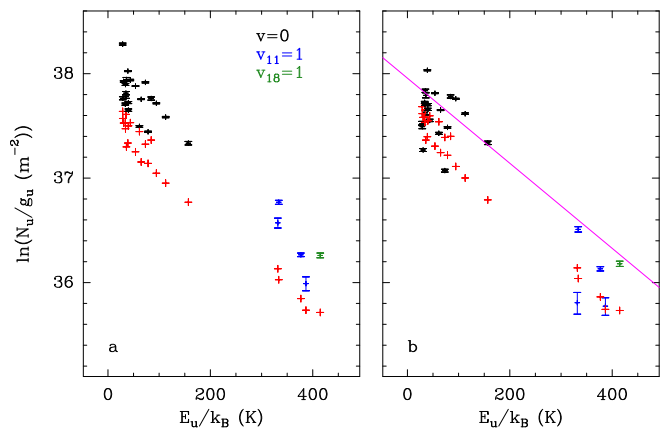}}}
		\caption{Population diagram of \ce{NH2CH2CN} toward Sgr~B2(N1S). The observed data points are shown in various colors (but not red) as indicated in the upper right corner of panel \textbf{a} while the synthetic populations are shown in red. No correction is applied in panel \textbf{a}. In panel \textbf{b}, the optical depth correction has been applied to both the observed and synthetic populations and the contamination by all other species included in the full model has been removed from the observed data points. The purple line is a linear fit to the observed populations (in linear-logarithmic space).}
		\label{f:popdiag_aan}
	\end{figure}

	Using the transitions that are not too much contaminated by emission from other species, we constructed a population diagram for aminoacetonitrile following the same method as \citet{Belloche16}. Figures~\ref{f:popdiag_aan}a and b show this diagram before and after correction for optical depth and contamination by emission of other species, respectively. The fit to the population diagram of Fig.~\ref{f:popdiag_aan}b yields a rotational temperature of $245 \pm 47$~K. As discussed in Sect.~4.4 of \citet{Belloche19}, there are several limitations to this population diagram, in particular because of the non-negligible level of background continuum emission that varies with frequency and angular resolution, and because of the residual contamination of the selected aminoacetonitrile transitions by emission from still unidentified species. These are the reasons why the fitted rotational temperature is uncertain, and its (purely) statistical uncertainty may be underestimated. For the model, we assumed a temperature of 200~K, which yields a good fit to the observed spectrum for both the vibrational ground state and the first two vibrationally excited states (see Figs.~\ref{f:spec_aan_ve0}--\ref{f:spec_aan_v18e1}).
	
	There is an inconsistency for one transition of $\varv_{11}=1$ at 109\,248~MHz where the synthetic spectrum largely overestimates the observed spectrum (see third panel of fifth row of Fig.~\ref{f:spec_aan_v11e1}). This may be due to a blend with absorption lines of $^{13}$\ce{CN} produced by translucent or diffuse clouds along the line of sight to Sgr~B2 against the strong continuum background of Sgr~B2(N1). The $F_1$=1-0 multiplet of the $N$=1-0, $J$=1.5-0.5 component of $^{13}$\ce{CN} has a rest frequency of $\sim$109\,218~MHz and is detected in absorption in the spectrum of Sgr~B2(N1S) at the velocity of Sgr~B2(N)'s envelope \citep[see the absorption feature in the second panel of the fifth row of Fig.~\ref{f:spec_aan_v11e1}, as well as][]{Thiel19b}. The frequency 109\,248~MHz corresponds to a velocity shift of $-82$~km~s$^{-1}$ with respect to the velocity of Sgr~B2(N1S), i.e. a systemic velocity of $-20$~km~s$^{-1}$ that corresponds to known absorbing clouds in the 4~kpc arm of the Galaxy \citep[][]{Thiel19a}. Therefore, the inconsistency at 109\,248~MHz does not invalidate our identification of $\varv_{11}=1$ emission toward Sgr~B2(N1S).
	
	We also searched for rotational transitions of aminoacetonitrile from within higher vibrational states, whose spectral predictions were prepared using the newly determined spectroscopic parameters for the $\varv_{17}=1$ state and the constants from \citet{kolesnikova2017rotational} for the $\varv_{11}=2$, $\varv_{11}=\varv_{18}=1$, and $\varv_{18}=2$ states.
	No transitions of $\varv_{17}=1$ and $\varv_{11}=2$ are unambiguously detected, but some of them contribute significantly to the signal detected with ALMA (see Figs~\ref{f:spec_aan_v17e1} and \ref{f:spec_aan_v11e2}, respectively). While this is not sufficient to claim a detection of these states, we included them in our full model to account for their contribution. 
	
	The next two vibrational states, $\varv_{11}=\varv_{18}=1$ and $\varv_{18}=2$, have a few rotational transitions with expected peak intensities at the level of 3--4$\sigma$ according to our LTE model, but they are unfortunately all fully blended with much stronger emission of other species. Therefore, we did not include these states in our full model.
	
	%---------------------------------------------------------------
	
	\section{Discussion}\label{sec:discussion}
	
	\subsection{Discussion of the spectroscopic results}
	\label{ss:discussion_spectro}
	
	The analysis carried out in the present paper represents a significant improvement in the spectroscopic knowledge of aminoacetonitrile.
	Prior to this work, the centrifugal analysis of the aminoacetonitrile spectrum was fairly extensive for the ground state only \citep{motoki2013submillimeter} and more limited for the excited states \citep{kolesnikova2017rotational,degli2017millimeter}.
	Here, this difference has been levelled off for two of the three lowest excited states, the exception being the $\varv_{11}=1$ states whose band is much weaker than the others.
	Probing energy levels with $J$ up to 80 and $K_a$ up to 25, we were able to remarkably extend the centrifugal analysis of aminoacetonitrile; the set of spectroscopic parameters now includes more reliable centrifugal distortion constants as well as additional centrifugal dependencies of the Coriolis term $G_a$. Furthermore, in this work we adopted a different choice of fitting parameters with respect to \citet{degli2017millimeter}. This results in a different determination of some spectroscopic constants, which is more pronounced for the purely $K_a$-dependent parameters (e.g. $A$, $D_K$, and so on).
	
	However, the most important result attained in our analysis is the determination of the vibrational energies of all the three excited states of aminoacetonitrile with relative uncertainties as low as $2 \times 10^{-8}$. It has to be noticed that our derived values differs by 2--6\,\wn from the energies reported in \citet{bak1975vibrational}, where the claimed accuracy was $\pm1$\,\wn.
	Larger discrepancies are found in comparison with low-level theoretical calculations \citep{bernstein2004infrared}, thus requiring improved levels of theory and approximations to achieve reliable results, as shown e.g. in \citet{melli2018rotational} for ethanimine.
	Last, we confidently confirm that the resonance effects between the $\varv_{11}=1$ and $\varv_{18}=1$ states are adequately accounted for and the inclusion of Coriolis interaction parameters is crucial to reproduce the spectrum of aminoacetonitrile. Although the treatment of resonances can be overcome by fitting ``effective'' spectroscopic parameters \citep{kolesnikova2017rotational}, this approach is not appropriate when spectral prediction must rely on frequency extrapolations of higher $J$ and $K_a$ transitions, which might be the case for COMs detected in hot cores.
	
	Our new spectral predictions should be reliable for astronomical identification of aminoacetonitrile from the microwave to the terahertz domain, for both the ground and vibrationally excited states.

	\subsection{Discussion of the astronomical results}
	\label{ss:discussion_astro}
	
	As reported in Sect.~\ref{ss:detection}, only four and one spectral lines of $\varv_{11}=1$ and $\varv_{18}=1$, respectively, are found to be little contaminated by emission of other species. While this would not be sufficient to claim the secure detection of a new molecule, the fact that these lines are well reproduced by our LTE model that uses the same consistent set of parameters as the vibrational ground state shows that LTE is a good assumption to describe the level populations of aminoacetonitrile in its vibrational states as well. This gives us strong confidence in the detection of both vibrational states toward Sgr~B2(N1S), also because our modelling procedure takes into account the contribution of all identified molecules and thus reduces the 
	risks of misassignments.
	
	The column density of aminoacetonitrile derived from our LTE modelling of the ReMoCA spectrum toward Sgr~B2(N1S), $1.1 \times 10^{17}$~cm$^{-2}$, is 3.9 times higher than the column density reported in \citet{Belloche08} for Sgr~B2(N) from observations with the IRAM 30~m telescope, although in both cases we assumed the same emission size. There are several reasons for this difference. First of all, \citet{Belloche08} did not account for the contribution of the vibrational partition function. At the temperature they assumed, 100~K, this contribution amounts to a factor 1.09 and, accounting for this, the column density  derived from the IRAM 30~m data becomes $3.0 \times 10^{16}$~cm$^{-2}$. Second, the column density of \citet{Belloche08} was computed at 100~K while here we used a temperature of 200~K. Assuming a temperature of 100~K like 
	\citet{Belloche08} would underestimate the intensities of the higher-energy transitions detected in the ReMoCA spectrum but we would obtain a column density of $\sim$$4.3 \times 10^{16}$~cm$^{-2}$ on the basis of the lower-energy ones. Finally, the column density derived here characterizes the position Sgr~B2(N1S) only, as probed with an angular resolution of 0.6$\arcsec$. At this scale, the velocity dispersion of the aminoacetonitrile emission is 5 km~s$^{-1}$, somewhat smaller than the velocity dispersion probed with the 30~m telescope (7~km~s$^{-1}$). Spatial variations across the hot core Sgr~B2(N1), completely covered by the single-dish beam, and calibration uncertainties could explain the remaining $\sim$30\% discrepancy.
	
	\subsection{Formation of aminoacetonitrile}
	
	The interstellar chemistry of aminoacetonitrile was first simulated by \citet{Belloche09}, who incorporated it into their gas-grain chemical network along with new chemistry for n-propyl cyanide, ethyl formate, and other species related to all three. The network they developed for aminoacetonitrile, as with other complex organics, relied on grain-surface/ice chemistry to produce the molecule from simpler, more abundant building blocks at low to intermediate temperatures, with the aminoacetonitrile produced on the grains ultimately being released into the gas phase at typical hot-core temperatures greater than $\sim$100~K. A selection of ion-molecule destruction routes for gas-phase aminoacetonitrile were also included.
	
	Building on this network to include glycine-related chemistry, \citet{Garrod13} employed a three-phase model of hot-core chemistry, in which the ice-surface and bulk were treated independently. This model included a selection of other updates and improvements to the treatment of surface/bulk chemistry; we therefore discuss the later results in preference, although in fact the general behavior of the aminoacetonitrile chemistry is not too dissimilar between the two models.
	
	The current understanding of aminoacetonitrile production, based on astrochemical kinetic models, therefore assumes production on grains. \citet{Garrod13} finds that it is predominantly formed through the addition of the radicals \ce{NH2} and \ce{CH2CN} within and upon the ice mantles, although another radical-addition process, between \ce{NH2CH2} and \ce{CN}, is also present in the network and makes a modest contribution. The radicals themselves are produced by direct photodissociation of stable, solid-phase species such as ammonia and methyl cyanide, or through the abstraction of hydrogen from those species by reactive and abundant radicals like \ce{OH}, which itself is formed through the photodissociation of water. This photodissociation is caused by the weak, secondary UV field induced by cosmic-ray collisions with gas-phase \ce{H2}. \citet{Garrod13} also discerned an increase in aminoacetonitrile production associated with the thermal desorption of methyl cyanide, whose gas-phase destruction routes included production of the \ce{CH2CN} radical; the latter species could produce aminoacetonitrile by re-accreting onto the grain/ice surface and there reacting with \ce{NH2}.
	
	Both chemical models produced fractional abundances with respect to \ce{H2} on the order of $10^{-8}$. Comparing the peak aminoacetonitrile abundance with that of methanol obtained in the medium warm-up timescale model of \citet{Garrod13} provides a ratio $\sim$$1.3 \times 10^{-3}$. Based on the column density data obtained in the present work, and using an observational methanol column density of $2.0 \times 10^{19}$ cm$^{-2}$ for Sgr B2(N1S) \citep[][]{Motiyenko20}, an observational ratio of $5.5 \times 10^{-3}$ is found. The models therefore appear roughly consistent with the observations, suggesting that photochemistry within the ices remains a plausible formation route for aminoacetonitrile.
	
	\section{Conclusions}\label{sec:concl}
	
	We presented a combined spectroscopic and observational study of vibrationally excited aminoacetonitrile, a species considered closely related to amino acids. The high-resolution far-infrared spectrum of aminoacetonitrile has been recorded for the first time using a synchrotron-based FT spectrometer. Three bands were observed in the frequency region between 100 and 500\,\wn, corresponding to the \ce{CCN} bending ($\nu_{11}$), \ce{NH2} torsion ($\nu_{18}$), and \ce{CH2}-\ce{NH2} torsion ($\nu_{17}$) modes.
	Then, rotational signatures from aminoacetonitrile in the ground and vibrationally excited states was searched for in the ReMoCA imaging spectral line survey of Sgr~B2(N). A number of 23 lines with lower energy levels ranging from 29 to 422\,K were detected in emission around 3\,mm toward the main hot core. The main results of this work can be summarized as follows:
	
	\begin{enumerate}
		
		\item The energies of the $\varv_{11}=1$, $\varv_{18}=1$, and  $\varv_{17}=1$ vibrationally excited states of aminoacetonitrile have been accurately determined from the analysis of almost 11\,000 ro-vibrational lines and around 2\,000 rotational data.
		
		\item Updated catalog entries of aminoacetonitrile in its ground and vibrationally excited states have been produced using the newly evaluated values of the vibration-rotation partition function together with the new set of spectroscopic constants.
		
		\item We report the interstellar detection of aminoacetonitrile in its vibrational states $\varv_{11}=1$ and $\varv_{18}=1$ toward the main hot core of Sgr~B2(N), with intensities consistent with expectations from an LTE model that fits well the emission of aminoacetonitrile in its vibrational ground state.
		
		\item The next two vibrational states, $\varv_{17}=1$ and $\varv_{11}=2$ contribute significantly to the observed spectrum, but cannot be identified securely due to blends with emission from other species. The next two states, $\varv_{11}=\varv_{18}=1$ and $\varv_{18}=2$ are not detected.
	\end{enumerate}
	This work demonstrates that a strong interplay between laboratory spectroscopy exploiting (sub)millimeter and synchrotron far-infrared techniques, and observational spectral surveys can lead to the detection of COMs in space and quantify their abundances. Future projects concerning the recording and analysis of infrared bands of aminoacetonitrile at frequencies higher than 500~cm$^{-1}$ are planned.
	
	%---------------------------------------------------------------
	
	\begin{acknowledgements}
		This work has been performed under the SOLEIL proposal \#20191573; we acknowledge the SOLEIL facility for provision of synchrotron radiation. We would like to thank the AILES beamline staff for their assistance and O.~Chitarra for her help during the proposal submission.
		The work at Bologna University was supported by RFO funds and MIUR (Project PRIN 2015: STARS in the CAOS, Grant Number 2015F59J3R).
		The work at SOLEIL was supported by the Programme National ``Physique et Chimie du Milieu Interstellaire'' (PCMI) of CNRS/INSU with INC/INP co-funded by CEA and CNES.
		The authors gratefully thanks the developers of the LWW program for providing us the installation package.
		This paper makes use of the following ALMA data: 
		ADS/JAO.ALMA\#2016.1.00074.S. 
		ALMA is a partnership of ESO (representing its member states), NSF (USA), and 
		NINS (Japan), together with NRC (Canada), NSC and ASIAA (Taiwan), and KASI 
		(Republic of Korea), in cooperation with the Republic of Chile. The Joint ALMA 
		Observatory is operated by ESO, AUI/NRAO, and NAOJ. The interferometric data 
		are available in the ALMA archive at https://almascience.eso.org/aq/.
		Part of this work has been carried out within the Collaborative
		Research Centre 956, sub-project B3, funded by the Deutsche
		Forschungsgemeinschaft (DFG) -- project ID 184018867.
		
	\end{acknowledgements}
	
	%---------------------------------------------------------------
	
	\bibliographystyle{aa}
	\bibliography{amino_fir}
	
	%---------------------------------------------------------------
	
	\begin{appendix}
		
		\label{appendix}
		\section{Complementary figures: Spectra}
		\label{a:spectra}
		
\begin{figure*}
			\centerline{\resizebox{0.82\hsize}{!}{\includegraphics[angle=0]{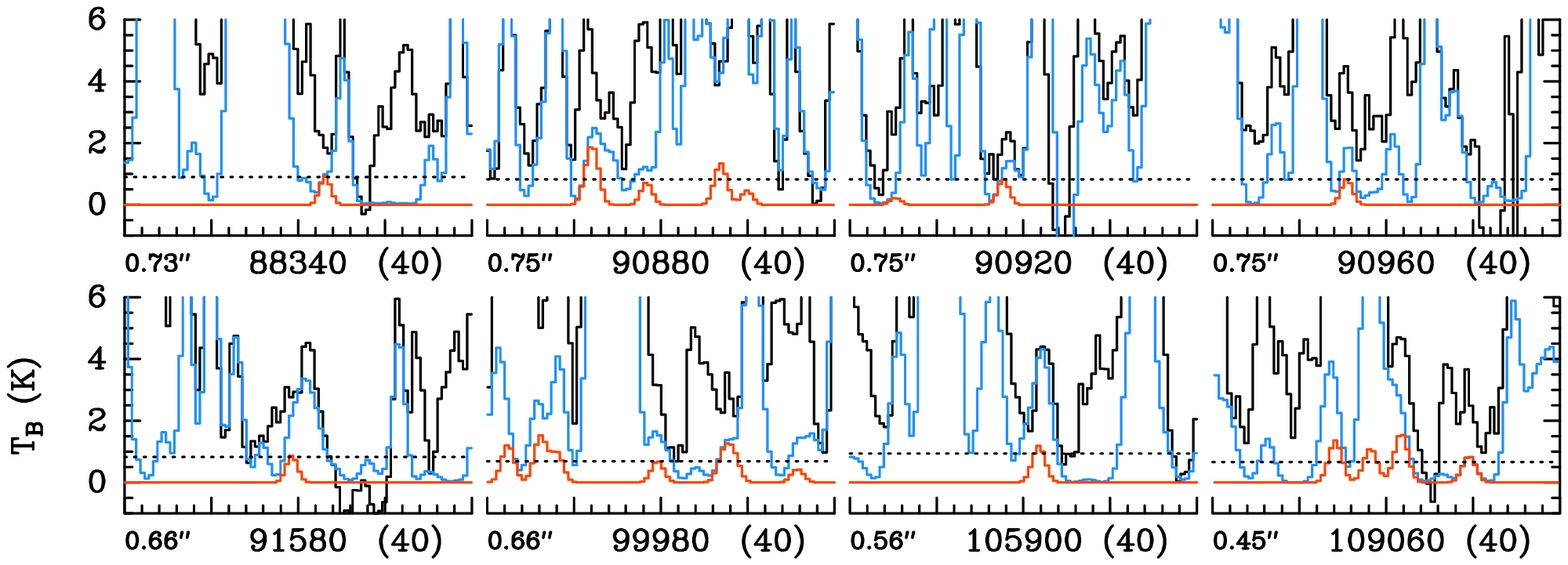}}}
			\caption{Same as Fig.~\ref{f:spec_aan_ve0} but for \ce{NH2CH2CN}, 
				$\varv_{17}=1$.}
			\label{f:spec_aan_v17e1}
		\end{figure*}

\begin{figure*}
			\centerline{\resizebox{0.82\hsize}{!}{\includegraphics[angle=0]{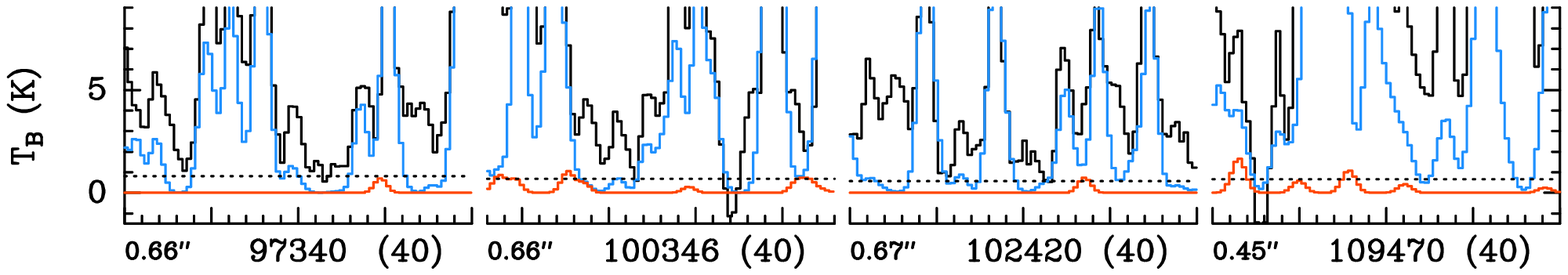}}}
			\caption{Same as Fig.~\ref{f:spec_aan_ve0} but for \ce{NH2CH2CN}, 
				$\varv_{11}=2$.
			}
			\label{f:spec_aan_v11e2}
		\end{figure*}

		Figures~\ref{f:spec_aan_v17e1} and \ref{f:spec_aan_v11e2} show rotational
		transitions of aminoacetonitrile from within its vibrationally excited states
		$\varv_{17}=1$ and $\varv_{11}=2$, respectively, that are covered by the ReMoCA 
		survey. Some of them significantly contribute to the signal detected toward 
		Sgr~B2(N1S). 
		
	\end{appendix}
	
\end{document}